\documentclass{article}
\usepackage{algorithm}
\usepackage{arxiv}
\usepackage{graphicx}
\usepackage[numbers]{natbib}
\usepackage[svgnames]{xcolor}
\usepackage{algorithm}
\usepackage{graphicx}
\usepackage{graphicx}
\usepackage{algcompatible}
\usepackage{amsmath}
\usepackage{amssymb}
\usepackage{caption}
\usepackage{balance}
\usepackage{listings}
\usepackage{color}
\usepackage{float}
\usepackage[utf8]{inputenc} % allow utf-8 input
\usepackage[T1]{fontenc}    % use 8-bit T1 fonts
\usepackage{hyperref}       % hyperlinks
\usepackage{url}            % simple URL typesetting
\usepackage{booktabs}       % professional-quality tables
\usepackage{amsfonts}       % blackboard math symbols
\usepackage{nicefrac}       % compact symbols for 1/2, etc.
\usepackage{microtype}      % microtypography
\usepackage{lipsum}		 
\usepackage{graphicx}
\usepackage{natbib}
\usepackage{doi}
\usepackage{multirow}
\usepackage{mathbbol}

\title{ Contrastive Environmental Sound Representation Learning }

\author{ \href{https://orcid.org/0000-0000-0000-0000}{\includegraphics[scale=0.06]{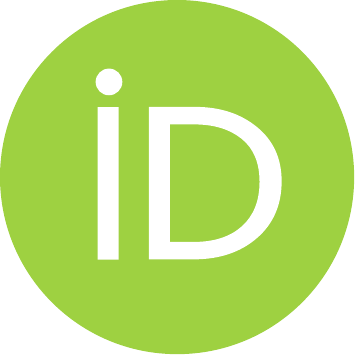}\hspace{1mm}Peter Ochieng}\thanks{} \\
	Department of Computer Science and Technology\\
		\texttt{po304@cam.ac.uk} \\
		University of Cambridge\\
	\And
	\href{https://orcid.org/0000-0000-0000-0000}{\includegraphics[scale=0.06]{orcid.pdf}\hspace{1mm}Dennis Kaburu} \\
	Department of Information Technology\\
	\texttt{dennis.kaburu@jkuat.ac.ke } \\
	Jomo Kenyatta University of Agriculture and Technology\\
}

\renewcommand{\shorttitle}{\textit{}}

\hypersetup{
pdftitle={A template for the arxiv style},
pdfsubject={q-bio.NC, q-bio.QM},
pdfauthor={Peter Ochieng, Dennis Kaburu},
pdfkeywords={First keyword, Second keyword, More},
}

\begin{document}
\maketitle

\begin{abstract}

\end{abstract}
Machine hearing of the environmental sound is one of the important issues in the audio recognition domain. It gives the machine the ability to discriminate between the different input sounds that guides its decision making. In this work we exploit the self-supervised contrastive technique and a shallow 1D CNN to extract the distinctive audio features (audio representations) without using any explicit annotations.We generate representations of a given audio using both its raw audio waveform and  spectrogram and  evaluate if  the proposed learner  is  agnostic to the type of audio input. We further use canonical correlation analysis (CCA) to fuse representations from the two types  of input of a given  audio  and demonstrate that the fused global feature results in robust representation of the audio signal as compared to the individual representations. The evaluation of  the proposed technique is done  on both ESC-50 and UrbanSound8K. The  results show that  the proposed technique is able to extract most features of the environmental audio and gives an improvement of 12.8\% and 0.9\%  on the ESC-50  and  UrbanSound8K datasets respectively.

% keywords can be removed
\keywords{Audio classification \and Environmental Sound\and Contrastive learning\and Machine hearing\and Unsupervised Learning}

\section{Introduction}
Human beings are able to hear environmental sounds and discriminate  between the different sounds. This is due to the fact that  they are equipped with auditory systems that capture sounds and extract meanings from them in a discriminative way \cite{clap2000}. These meanings are necessary in informing how humans make decisions and  respond or behave based on the meaning of the sound extracted. Enabling machines to have sensing capabilities such as those of humans e.g. vision, hearing, touch, smell and taste is part of the goal of machine learning \cite{revaudio200}. To this end a number of machine learning  techniques have been developed that focus on giving  machines  auditory skills similar to those of human beings\cite{tool2000}, \cite{tool2001},\cite{tool2002}. This challenging problem is referred to as machine hearing \cite{revaudio200}. A given hearing machine will be faced with a wide variety of  sounds that they are required to perform an in-depth analysis of to extract their appropriate features that make them distinct. The process of extracting  distinctive features of an audio signal is referred to as feature extraction. Accurate and precise  feature extraction is crucial in machine hearing to guarantee the success of machine hearing applications. The features extracted from the sounds can then be used  in different tasks such as audio classification, detection, retrieval etc.
Recently research on making machines to hear  environmental sound and classify them correctly has gained traction \cite{TFCNN100}, \cite{pyramid100},\cite{Piczac100},\cite{Li2018}. Environmental sound classification (ESC), is geared to make machines identify environmental sounds  such as  siren, bird chirping, car horn etc and distinguish the sounds. 
Compared  to music and speech audio, environmental sounds have
a number of distinctive  characteristics that present additional challenges to the machine hearing devices. Some of distinctive  features  between speech and audio vs environmental sounds include;
\begin{enumerate}
\item  In speech and music audio signals,  their respective phonemes and musical notes are combined so that the hearing human or machine can obtain a sequence of meanings such that they transmit a particular semantic message. However,  environmental sounds do not follow any predefined grammar and the semantic sequences remain unclear \cite{revaudio200}
\item Both  speech and music sounds are constructed  from a limited dictionary of phonemes and notes  respectively. On the other hand, environmental sounds are theoretically composed of infinite sounds from the environment since any occurring sound in the environment may be included in this category.
\item The environmental sounds depict a larger complexity of the spectrum  in the frequency domain as compared to the music and speech sounds \cite{revaudio200}.
\end{enumerate}
Based on these key differences, a number of  state of the art techniques \cite{pyramid100},\cite{Piczac100},\cite{Li2018} have been developed that focus solely on  environmental noise feature extraction and  classification. Most of the tools first design  techniques that extract audio signal features then  use the features to perform classification. All  of the tools reviewed have applied the supervised machine learning technique as the main technique of extracting the audio features. One of the  core objectives of deep learning is to learn useful representations of input data without annotated dataset \cite{con2000}. In the recent past, self-supervised methods have shown great success in domains such as computer vision \cite{visual200}, natural language processing \cite{electra2000} and speech recognition \cite{wav2vec200}. These self-supervised methods learn to extract distinctive features from the input dataset  without explicit annotations by recasting  the unsupervised representation learning problem into a supervised learning problem. Motivated by this, we also introduce the self-supervised contrastive technique into the environmental sounds domain. We specifically use the contrastive technique to learn how to extract representations of the different environmental sounds.
Concretely we make the following contributions;
\begin{enumerate}
\item Demonstrate that contrastive self-supervised technique can be used to extract  robust feature representations of different environmental sounds.
\item We adopt a shallow 1D CNN model to extract features of an environmental sound and evaluate  the effect of increasing  the depth of the  1D CNN model in the ESC accuracy.
\item We adopt mini-batch balancing during model training and evaluate its effect on  the classification accuracy of minority classes.
\item We exploit canonical correlation analysis to fuse two feature representations from raw normalised  waveform and spectrogram  inputs and evaluate the accuracy of the merged representation in the ESC.
\end{enumerate}
\section{Related Work}
Here, we review the different techniques that have been used in the classification of the environmental sounds. In each technique, we discuss, the type of input, the machine learning technique model for capturing features and how the model was trained  i.e. supervised or unsupervised. SB\_CNN \cite{Salamon2017} converts  raw audio into log-scaled Mel spectrogram representations. It then extracts 2D frame patches from the spectrograms which are then processed by a convolution neural network (CNN) model. The CNN model is connected to a 2 layer fully connected MLP with Softmax activation function in the output layer. The model is trained via a supervised training  to capture the  audio representation. The tool proposed in \cite{Piczac100} first  extracts log-scaled Mel spectrogram of the  audio then splits the spectrogram into two overlapping segments which act as the input of the proposed CNN model. It then performs  a cross validation supervised training to learn audio representations. Pyramid-Combined CNN \cite{pyramid100} first converts input audio waveforms into a spectrogram. It then uses average pooling and normalisation to  obtain the first transformed spectrogram. It again applies the same process of average pooling and normalisation on the first transformed spectrogram to obtain the second transformed spectrogram.  Combined with the original untransformed  spectrogram, three spectrograms are obtained. It then uses  learned weights of the pre-trained CNN models to extract  deep features from the three spectrograms. The obtained features are then fused into  a single representation where  normalisation is applied.  Relief method is then applied on the fused representation to  select the most discriminative features.  Finally, an SVM classifier is used to determine the class labels of the input waveforms. In ACRNN \cite{ACRNN100} first converts an audio to its spectrogram form, then uses a CNN model to extract high level feature representation of the spectrogram. The established features are then fed into a bidirectional gated recurrent neural network to learn the temporal correlation information. This is then passed to a fully connected layer for classification.  SoundNet \cite{Aytar2016} uses transfer learning technique . They leverage unlabeled video to learn a representation of sound. The tool uses CNN model to achieve a transfer learning between unlabeled video and raw waveform. From the features extracted from the CNN, a linear SVM is built for classification. The tool in \cite{verydeep100} uses raw waveform as the input to a 1D CNN. The 1D CNN is used to capture the features of the waveform.  It experiments with different depths of 1D CNN to compare performance differences. DS-CNN \cite{Li2018} uses both raw -waveform and spectrogram as its input. It uses two CNN models where one captures the features of the raw waveform and the other captures the features of the spectrogram. The features representation from the two inputs are combined and classification done based on the combined features. Table 1 summarises  the tools.
 \begin{table}[ht]
\centering
\caption{Summary of the state of the art tools for environmental sound classification}
 \begin{tabular}{ |c |c |c|c|}
   \hline
   \textbf{Tool} &\textbf{Technique}&\textbf{Type of Input}  &\textbf{Type of Training}  \\ \hline
   SB\_CNN \cite{Salamon2017} &2D CNN &spectrogram& Supervised \\ \hline
   Piczak-CNN \cite{Piczac100}&2D CNN& spectrogram& Supervised \\ \hline
    ACRNN \cite{ACRNN100}&2D CNN+GRU& spectrogram& Supervised \\ \hline
         SoundNet100\cite{Aytar2016} &2D CNN& raw waveform& Supervised \\ \hline
           \cite{verydeep100} &1D CNN& raw waveform& Supervised \\ \hline
             DS-CNN\cite{Li2018}&2 D CNN& raw waveform+spectrogram& Supervised \\ \hline
               Pyramid CNN \cite{pyramid100}& 2 D CNN& spectrogram& Supervised \\ \hline
TFCNN  \cite{TFCNN100}& 2 D CNN& spectrogram& Supervised \\ \hline
     \end{tabular}
     \end{table}

\section{Model}
Our model exploits the unsupervised contrastive learning  \cite{contrastive200},\cite{contrastive3000},\cite{contrastive400} to establish audio representations. The model shown in figure 1 is  composed of two key parts;
\begin{enumerate}
\item Input preprocessing.
\item Feature encoder
\end{enumerate}
\begin{figure}[ht]
	\centering
\includegraphics[scale=0.45,angle=0]{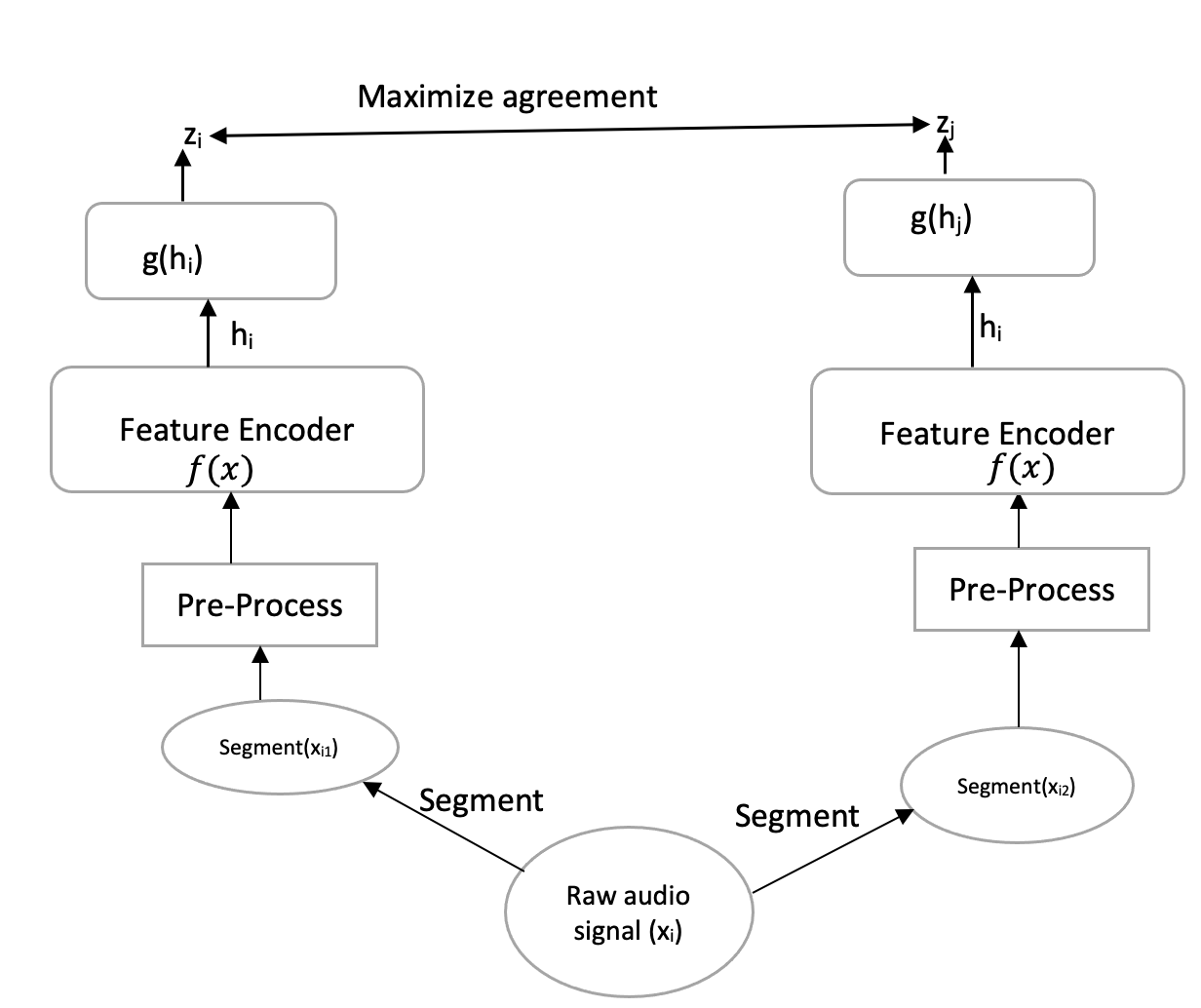}
		\caption{Overall structure of the proposed self-supervised learning for  audio representation of environmental sound }
	\end{figure}
\subsection{Input  Pre-processing}
The feature encoder  proposed in this work  is able to accept two types of inputs.   The first type is the normalised waveform similar to the one adopted in \cite{wav2vec200}. Here, we first segment the raw waveform of length $t$ into two equal parts i.e. each waveform segment has  a length $t/2$. The  waveform  segments'  values are then normalised  to zero mean and unit variance. The normalised waveform acts as the first type of  input to the feature encoder can accept.  The second type of input that can be accepted by the feature encoder is the spectrogram patches. Here, we  segment raw audio of time $t $ into two segments  each of length $t/2$. From each segment we extract a sequence of 128-dimension log Mel spectrogram features  computed using a 25 ms Hamming window with  10 ms stride  yielding  a spectrogram  $X$ of size ${128\times 100t}$.  We then randomly select a number of rectangular patches  in time from the full Mel spectrogram to act as the input to the feature encoder.  Each $x_i$ patch has the shape  $128\times 100t$ where $t$ is the length of the patch.
  \subsection{Feature Encoder}
For the feature encoder,  we use a  1D convolution neural network (CNN). The 1D CNN has
been used successfully in other domains such as  in fault detection \cite{fault2000} and  patient electrocardiogram classification \cite{real200}.  It was utilised in these domains since it  has a  low computational complexity and it is able to  extract features of   dataset without any predetermined transformation. Since the proposed feature encoder can accept   both raw waveform  and spectrogram  as its input, 1D CNN  was considered ideal.  The main unique  architectural design of the 1D CNN is that each CNN layer  is equipped to perform both convolution and subsampling \cite{real200} \cite{fault2000}( see fig 2). To compute the input of a neuron $k$ at layer $l$, the output maps of the previous layer $l-1$ are convolved with their respective kernels and then summed to obtain  $x_k^l$  the input of neuron $k$ at layer $l$ (eqn 1).
 \begin{equation}
x_k^l=b_k^l+\sum_{i=1}^{N} Conv1D(w_{ik},s_i^{l-1})
\end{equation}
where  $w_{ik}^{l-1}$ is the weight connecting $i^{th}$ neuron at layer $l-1$ and $k$ neuron in  layer $l$.  $s_i^{l-1}$ is the output of the $i$ neuron at layer $l-1$, $b_k^l$ is the bias of the $k^{th}$ neuron at layer $l$. To compute the output $s_k^l$of  the $k^{th}$ neuron at layer $l$, $x_k^l$  processed via  a non-linear function $f$ ( eqn 2) then sub-sampling is applied (eqn 3).
\begin{equation}
y_k^l=f(x_k^l)
\end{equation}
\begin{equation}
s_k^l=Subsampling(y_k^l)
\end{equation}
   \begin{figure}[ht]
     \centering
\includegraphics[scale=0.5,angle=0]{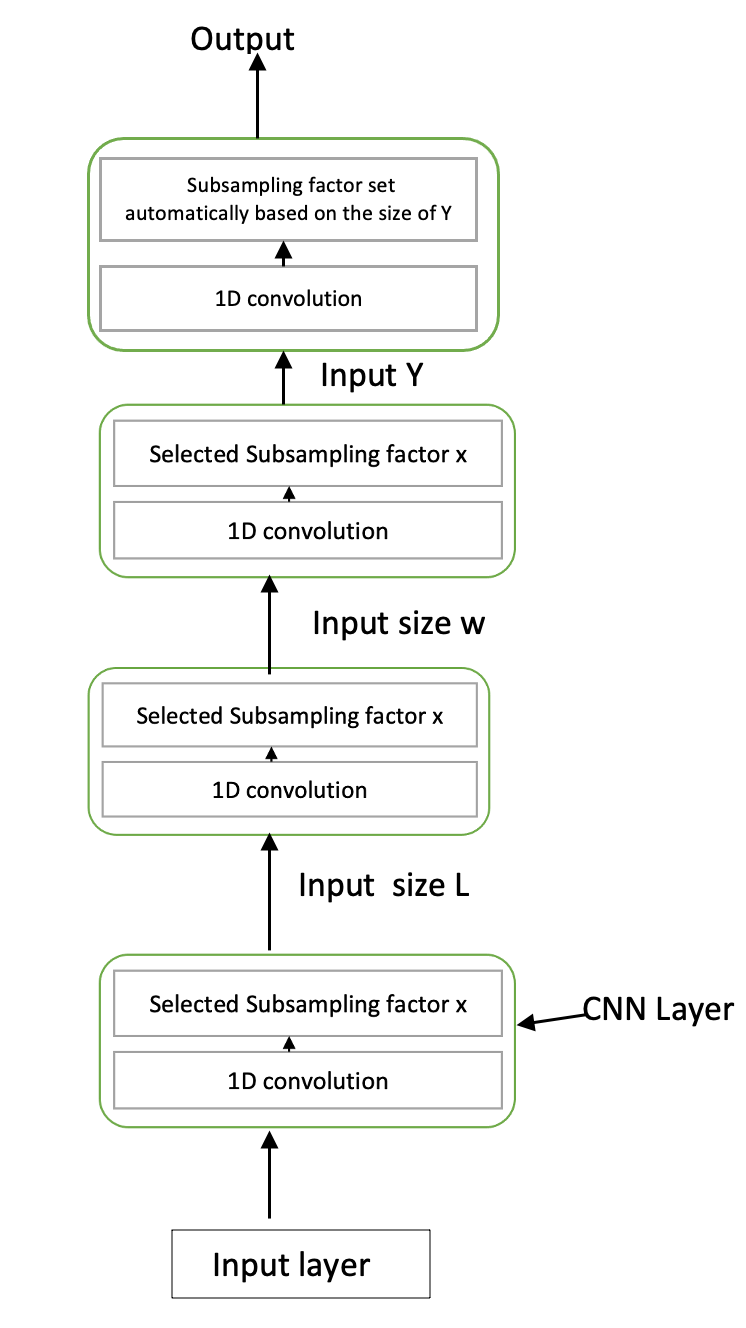}
		\caption{1D CNN with four CNN layers }
	\end{figure}
	
 The 1D CNN architecture is also  adaptive and allows for any number of hidden layers. This is enabled by the fact that the last sub-sampling factor is dynamically set to be equal to the size of  its input map.
\section{Model Training with normalised waveform as the input}
The self-supervised contrastive framework for generating environmental sound representation is based on the idea that two  audio waveform segments $x_i$ and $x_j$ generated  from a given waveform $X$ are likely to contain some level of overlapping  information. This is based on our  analysis of UrbanSound8K \cite{urbandataset200} dataset  where the majority of the waveform visualisation has a repetitive nature hence exhibiting high periodicity ( see fig 3 for some visualisations of  selected raw waveforms). During training we seek to learn environmental sound audio representations by setting up a contrastive task  $\mathcal{C}$ which requires that a segment $s_i$ should be able to identify its sibling  from a set of other segments. To compute the loss $\mathcal{L}$,  we sample a random mini-batch of $M$  examples of raw waveforms and segment each into two equal parts resulting in $2M$ raw waveforms each of  size $t/2$. Each of the $2M$ raw waveforms are then normalised to zero mean and unit variance. The normalised waveforms serve as the input of the feature encoder. The feature encoder generates waveform representation $h$ which is then projected to an MLP network which produces final representation $z$ where contrastive loss is applied.  Denoting as $z_i$ and $z_i^\prime$ the two segment representations of  audio sounds of the $i^{th}$ audio input, the contrastive loss is defined  according to equation 4. 
\begin{equation}
\mathcal{L}=\frac{1}{M} \sum_{i=1}^{i=M} L_i
\end{equation}
and
\begin{equation}
L_i=\log \frac{\exp( \nicefrac{sim(z_i,z_i^\prime)}{\tau})}{\sum_{k=1}^{i=2M}  \mathbb{1}_{[k\neq i]}  \exp(\nicefrac{sim(z_i,z_k)}{\tau})
}
\end{equation}
where $sim(u,v)=\frac{u^Tv}{||u||||v||}$
\begin{figure}[ht]
	\centering
\includegraphics[scale=0.45,angle=0]{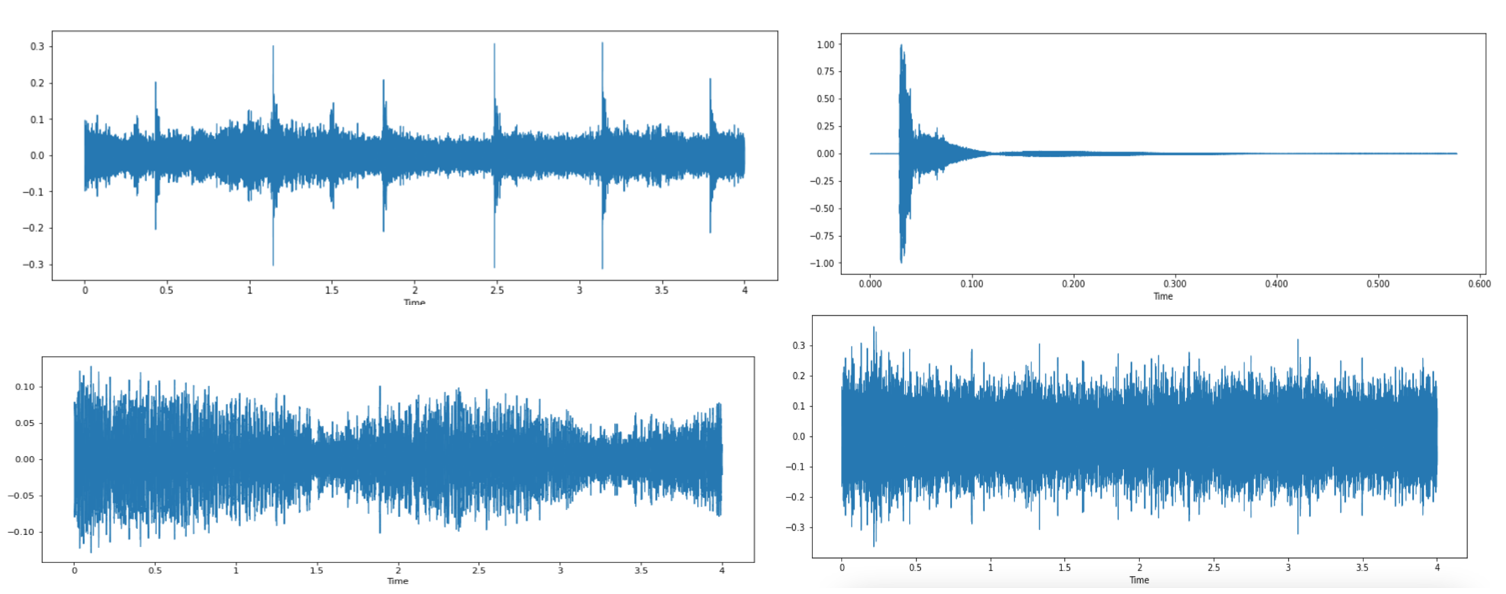}
		\caption{Raw waveform of car horn, gun shot, street music and air conditioner respectively. }
	\end{figure}
\section{Model Training with Spectrogram as input}
Here,  we sample a random mini-batch of $M$  examples of raw waveforms and segment each into two equal parts resulting in $2M$ raw waveforms each of  size $t/2$. From each segment we extract  a sequence of 128-dimension log Mel spectrogram features as described in section 3.1. From the full Mel spectrogram of a segment, we then randomly select rectangular patches in time.  Each patch has a fixed duration of 1.5 s resulting in a rectangular patch of shape $P\in R^{128\times 150}$. The patches are then mapped to 1D  vector embedding  of size 768  through a trainable non-linear projection( see eqn 6) introduced just before the feature encoder. The embeddings are  then projected to the feature encoder and finally to the MLP where contrastive learning is applied according to equation 4 and 5.
\begin{equation}
k=f(W(P_1^TE+P_2^TE+\cdots+P_N^TE))
\end{equation}
where $E\in R^{128\times 768}$ and $N$ is the total number of randomly patches selected per Mel spectrogram of a segment and $W\in R^{1\times 150}$ and $f$ is a nonlinear function.

\section{Feature Fusion of  waveform and spectrogram representation}
Here both  the normalized raw waveform and  the extracted patches of spectrogram are fed into two trained  contrastive learners as shown in figure 1. The two feature representations generated by   the two contrastive learners are fused using canonical correlation analysis (CCA) \cite{Hardoon2003}, \cite{fusion2000} into a single global representation. We begin by giving a brief review of key intuition behind CCA. Consider two multivariate random vector ($\textbf{x}, \textbf{y})$. The idea behind CCA is  to find the basis vectors $\beta$ and $\alpha$ for two sets of variables $\textbf{x}$ and $\textbf{y}$ such that the correlation between the projections of  $\textbf{x}$ and $\textbf{y}$  onto the basis vectors $\beta$ and $\alpha$ i.e. $k_1=\beta^T x$ and $w_1=\alpha^T x$ are mutually maximized. If  $S_x=\{x_1,x_2,\cdots c_n\}$ and $S_y=\{y_1,y_2,\cdots y_n\}$,  CCA seeks to define a new direction  of $S_x$  by choosing a new direction $\beta$  and projecting the vector $S_x$ onto that direction i.e $S_x\beta=(\langle \beta,x_1 \rangle,\langle \beta,x_2 \rangle,\cdots,\langle \beta ,x_n\rangle)$. Similarly,  $S_y$ can be projected to a new directions by choosing a direction $\alpha$ i.e.
$S_y \alpha=(\langle \alpha,y_1 \rangle,\langle \alpha,y_2 \rangle,\cdots,\langle \alpha,y_n\rangle)$. The first task therefore is to select $\beta$ and $\alpha$ that maximise the correlation between the two vectors according to
\begin{equation*}
\sigma= \max_{\alpha,\beta} corr(S_x\beta S_y\alpha)=\max_{\alpha,\beta} \frac{\langle S_x\beta,S_y\alpha\rangle}{||S_x\beta||||S_y\alpha||}
\end{equation*}
Defining the empirical expectation  of a function $f(\textbf{x,y})$ as
\begin{equation*}
E[f(x,y)]=\frac{1}{m}\sum_{i=1}^{i=m} f(\textbf{x,y})
\end{equation*}
the expression of $\sigma$ can be rewritten as
\begin{equation*}
\sigma= \max_{\alpha,\beta} \frac{E[\langle \beta,\textbf{x}\rangle \langle \alpha, \textbf{y}\rangle]}{\sqrt{E[\langle \beta,\textbf{x}\rangle^2] E[\langle \alpha,\textbf{y}\rangle]^2]}}
\end{equation*}
\begin{equation*}
= \max_{\alpha,\beta} \frac{E[\langle \beta,\textbf{x}\rangle \langle \alpha, \textbf{y}\rangle]}{\sqrt{E[\beta^\prime \textbf{xx}^\prime \beta] E[\alpha^\prime\textbf{yy}^\prime \alpha)]}}
\end{equation*}
\begin{equation*}
= \max_{\alpha,\beta} \frac{\beta^\prime E[\textbf{xy}^\prime]\alpha^\prime}{\sqrt{\beta^\prime_x E[\textbf{xx}^\prime] \beta \alpha^\prime E[\textbf{yy}^\prime] \alpha)}}
\end{equation*}
Note that  the covariance matrix of (x,y) as
\begin{equation*}
cov(x,y)=E \left[ \begin{bmatrix}
          \textbf{x} \\
          \textbf{y} \\
       \end{bmatrix} \begin{bmatrix}
          \textbf{x} \\
          \textbf{y}\\
       \end{bmatrix}^\prime \right]=\begin{bmatrix}
          C_{\textbf{xx}} &  C_{\textbf{xy}} \\
          C_{\textbf{yx}} &   C_{\textbf{yy}}
        \end{bmatrix}
\end{equation*}
$cov(\textbf{x,y})$ matrix is composed of within set covariance $C_{\textbf{xx}}$ and $C_{\textbf{yy}}$ and the between the set covariance $C_{\textbf{xy}}=C_{\textbf{yx}}^\prime$
based on this, $\sigma$ can be rewritten as
\begin{equation}
\sigma=\max_{\alpha,\beta} \frac{\beta^\prime C_{\textbf{xy}} \alpha^\prime}{\sqrt{w^\prime_x C_{\textbf{xx}} \beta\alpha^\prime C_{\textbf{yy}} \alpha^\prime}}
\end{equation}
therefore canonical correlation is maximised based on the values of the basis vectors $\beta$ and $\alpha$
\par In this work, Given the two feature vectors $Z$  and $R$ , we exploit  CCA  to fuse the two features into a single representation that only keeps discriminant information and reduces redundancy between the two representations as much as possible. To fuse the two representations  we follow the following steps \cite{fusion2000};
\begin{figure}[ht]
	\centering
\includegraphics[scale=0.45,angle=0]{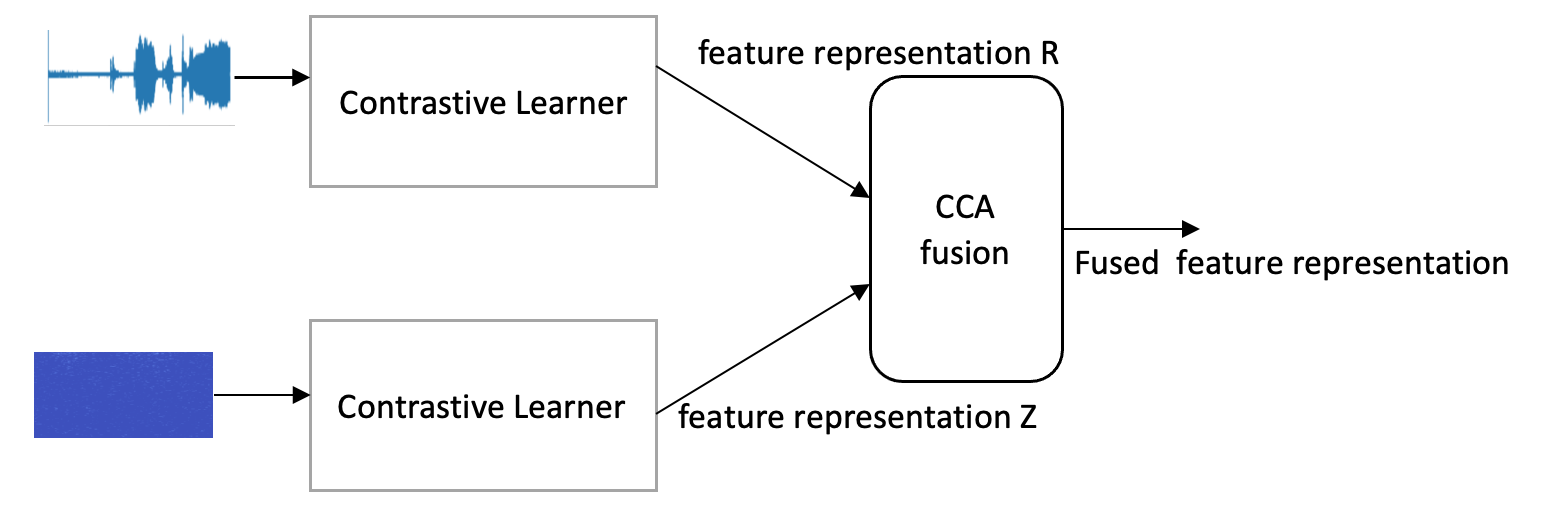}
		\caption{Fusing representation from different input types. }
	\end{figure}
First use the trained contrastive learners to extract two feature vectors representation of the raw waveform and the spectrogram (see fig 4). We then perform the canonical correlation analysis (CCA) between the two representations  for fusion as follows;
\begin{enumerate}
\item Compute  the covariance matrices $S_{rr}$,$S_{zz}$ and     $S_{rz}$ representing covariance matrix of $R$, $Z$ and between  $Z$ and $R$ respectively.
\item Compute  $G1=S_{rr}^{-1/2}S_{rz}S_{zz}^{-1}S_{zr}S_{rr}^{-1/2}$, $G2=S_{zz}^{-1/2}S_{zr}S_{rr}^{-1}S_{rz}S_{zz}^{-1/2}$ and establishing their non-zero eigenvalues $\lambda_1^2\geq\lambda_2^2\geq\cdots\geq\lambda_r^2$ and corresponding orthonormal eigenvectors $u_i$ and $v_i$  with $ 1\leq  i\leq r$
\item Compute iteratively the canonical projection vectors (CPV)  $\alpha_i$  and $\beta_i$ where $ 1\leq  i\leq r$. $\alpha_i$ and $\beta_i$ are computed such that they maximise the correlation between the projections $a_i=\alpha_i^TR$ and $b_i=\alpha_i^TZ$. The projections $a_i$ and $b_i$ are referred to as a pair of canonical variates. 
\item  Select the first $d$ pair of projections such that $R^*=(\alpha_1^Tz,\alpha_2^Tz,\cdots,\alpha_d^Tz)=(\alpha_1,\alpha_2,\cdots,\alpha_d)^T z=W_z^Tz$
and $Z^*=(\beta_1^Tr,\beta_2^Tr,\cdots,\beta_d^Tr)=(\beta_1,\beta_2,\cdots,\beta_d)^T z=W_r^Tr$.
\item Compute feature fusion as $T=X^*+Y^*$
\end{enumerate}
\section{Experimental Setup}
\subsection{Contrastive Training Details}
The 1D CNN was implemented with 4 convolution layers. For the three hidden layers we used 64,32 and 16 neurons respectively. The  kernel sizes used were  1,9,15 and 4 for the first, second, third and fourth CNN layers. We also use a sub-sampling value of 4 for the first three layers and the last CNN layer is set to 5 automatically. The output of the last CNN layer is fed into a 2 layer MLP for the final projection of the audio features to the 128-D representations. We use the Adam optimizer for the optimization, batch size of 64, learning rate of 0.001 and training steps of 400. ReLu activation is used in all layers. Further,  a dropout is  added to the fully connected layer with a probability of 0.5.
\subsection{Dataset}
We exploited  two datasets to evaluate the model proposed in this work. The first dataset is $ESC-50$ \cite{dataset2000} which has  50 unique classes of audio data. Each class has   40 audio data with a length of 5 s.  The $ESC-50$ dataset is balanced i.e. each class contains 40 audio data. The second dataset is the UrbanSound8K \cite{urbandataset200}  that contains annotated audio sounds collected from the urban setting. It contains  10 classes ( see table 1). The audio clips have varying lengths with the longest being 4 seconds. Further this dataset is class imbalanced i.e. some classes contain instances that are less than 1000. The distribution is shown in table 2.

\begin{table}[H]
\centering
\caption{Number of audio per class in UrbanSound8K dataset}
 \begin{tabular}{ |c | c | }
   \hline
   \textbf{Class} & \textbf{Number of instances} \\ \hline
   Street Music & 1000 \\ \hline
   Siren & 929 \\\hline
   Gun shot & 374 \\\hline
    Engine idling & 1000\\\hline
     drilling & 1000\\\hline
     dog bark & 1000\\\hline
      children playing  & 1000\\\hline
        car horn & 429\\\hline
         Air conditioner & 1000\\\hline
         \textbf{Total}&8732\\\hline
                 \end{tabular}
 \end{table}
\subsection{Mini-batch balancing}
The UrbanSound8K is unbalanced, therefore training the model with this unbalanced dataset may generate a model that shows weak generalisation ability for the minority classes. To mitigate against this, we adopt the balanced mini-batch training \cite{balanced2000}. We allow for an overlap selection of minority samples within the same epoch. Further, we restrict the number of samples from each class in a given mini-batch to be equal to the batch size divided by the number of classes.
\subsection{Evaluation objectives}
We conducted a number of experiments with a goal to establish if;
\begin{enumerate}
\item  The proposed method generates  environmental audio representation that results in accurate  classifications of the environmental sounds.
\item The input type significantly affect the quality of representations generated by the proposed method.
\item Increasing the depth of 1D CNN improves the quality of representations generated by the proposed model.
\item Mini-batch balancing has any effect in boosting the classifications of minority classes.
\end{enumerate}

\subsection{What is the quality of  audio representation generated by the proposed model }
The first experiment is to evaluate the ability  of representations generated by the proposed technique to capture the features of the audio dataset. To do this,  we set up a classification task by adding a 1  MLP  layer of size 10 for the UrbanSound8K  and 50 for  $ESC-50$ dataset on top of a trained contrastive learner. Our goal is to  establish the ability of the representations generated by the learner in classifying the environmental sounds correctly. We  evaluate the three possible configurations;
\begin{enumerate}
\item When the input to the contrastive learner is normalised raw waveform.
\item When the input to the contrastive learner is the spectrogram patches.
\item When there is fusion of both raw waveform and the spectrogram features.
\end{enumerate}
The results are shown in table 3.
\begin{table}[ht]
\centering
\caption{Performance of the three configurations proposed in this work.}
 \begin{tabular}{ |c | c | c|}
   \hline
   \textbf{Technique} & ESC-50 & UrbanSound8K \\ \hline
  
       Contrastive (raw waveform) & 94.42& 95.1 \\\hline
     Contrastive (spectrogram patches) & 94.9 & 95.43\\\hline
          Contrastive (fusion)  & 96.2 & 97.1\\\hline
   \end{tabular}
\end{table}
From the results in table 3, when raw waveform is the input, the model reports the lowest accuracy as compared to the other two. However,  the margin of results between  contrastive (raw waveform)  and contrastive (spectrogram patches) is marginal, signalling the ability of 1D CNN to capture the features of both untransformed and transformed waveforms well. The fused features between the  representations of raw waveform and spectrogram gives superior results demonstrating the need to capture feature representation from both the input types.
\section{What is the effect of  CNN depth on the quality of representations generated  }
Here, we experimented  with  different number of 1D CNN layers in the feature encoder  to establish the optimum depth of the 1D CNN model that produces audio representations that capture more accurate features of an audio input. We used a 1D CNN with 4,6,8 and 10 layers. The  number of neurons used in each layer is shown in table 4.  We mainly use  1,9,15 and 6 as the kernel sizes  and sub-sampling factors of 4 and 2.
\begin{table}[H]
\centering
\caption{ No. of neurons used in a given layer}
 \begin{tabular}{ |c | c | }
   \hline
   \textbf{Layer} & \textbf{Number of Neurons} \\ \hline
   2& 64 \\ \hline
   3 & 32 \\\hline
   4 & 16 \\\hline
    5& 64\\\hline
     6 & 32\\\hline
     7 & 16\\\hline
      8  & 62\\\hline
        9& 32\\\hline
         10& 16\\\hline
          \end{tabular}
          \end{table}
        
The results of the experiments are shown in table 5. From the results,  there is no definite trend with the varying number of layers. An increase in the depth of the 1D CNN does not offer significant benefit in terms of classification accuracy in all the model configurations. This may point to the ability of the 1D CNN to capture most features even when a shallow depth is used.
\begin{table}[H]
\centering
\caption{ Effects of varying CNN depth on the performance of the proposed model}
\begin{tabular}{|l|l|l|l|l|}
\hline

   & \textbf{Layers}& \textbf{ESC-50} &\textbf{UrbanSound8K}  \\ \hline
\multirow{4}{*}{\textbf{Contrastive(raw waveform)}} & 4 & 94.42 &95.1   \\\cline{2-4}
     & 6 & 94.67 &  95.04   \\ \cline{2-4}
  & 8 & 94.30 & 95.12    \\ \cline{2-4}
   & 10 & 94.16 & 95.28    \\ \hline
\multirow{4}{*}{\textbf{Contrastive(Spectrogram patches)}} & 4 & 94.9 & 95.43  \\ \cline{2-4}
 & 6 & 95.56 & 95.49  \\ \cline{2-4}
  & 8 &  94.8 & 95.83  \\ \cline{2-4}
  &10& 94.94&  95.92   \\ \hline
\multirow{4}{*}{\textbf{Contrastive(fusion) }} & 4& 96.2 & 97.1 \\ \cline{2-4}
  & 6 & 96.03 &97.41   \\ \cline{2-4}
 & 8 & 96.24 &  97.04   \\ \cline{2-4}
& 10 & 96.76 & 97.71    \\ \hline
\end{tabular}
\end{table}

\subsection{What is the effect of mini-batch balancing on the class prediction}
Here we evaluate if mini-balancing improves the classification of the minority classes in the UrbanSound8K. The evaluation uses  the contrastive(Spectrogram) model configuration. We first perform training when the mini-batch balancing technique has  not been applied during training. The results are shown in the confusion matrix figure 5. From the results the minority classes gunshot  and car horn have the lowest classification accuracy of 85\% and 89\% respectively. This shows that the model does not see enough instances  of the two classes to effectively capture their features. Figure 6 shows the confusion results when mini-batch balancing has been applied during training.
\begin{figure}[ht]
	\centering
\includegraphics[scale=0.4,angle=0]{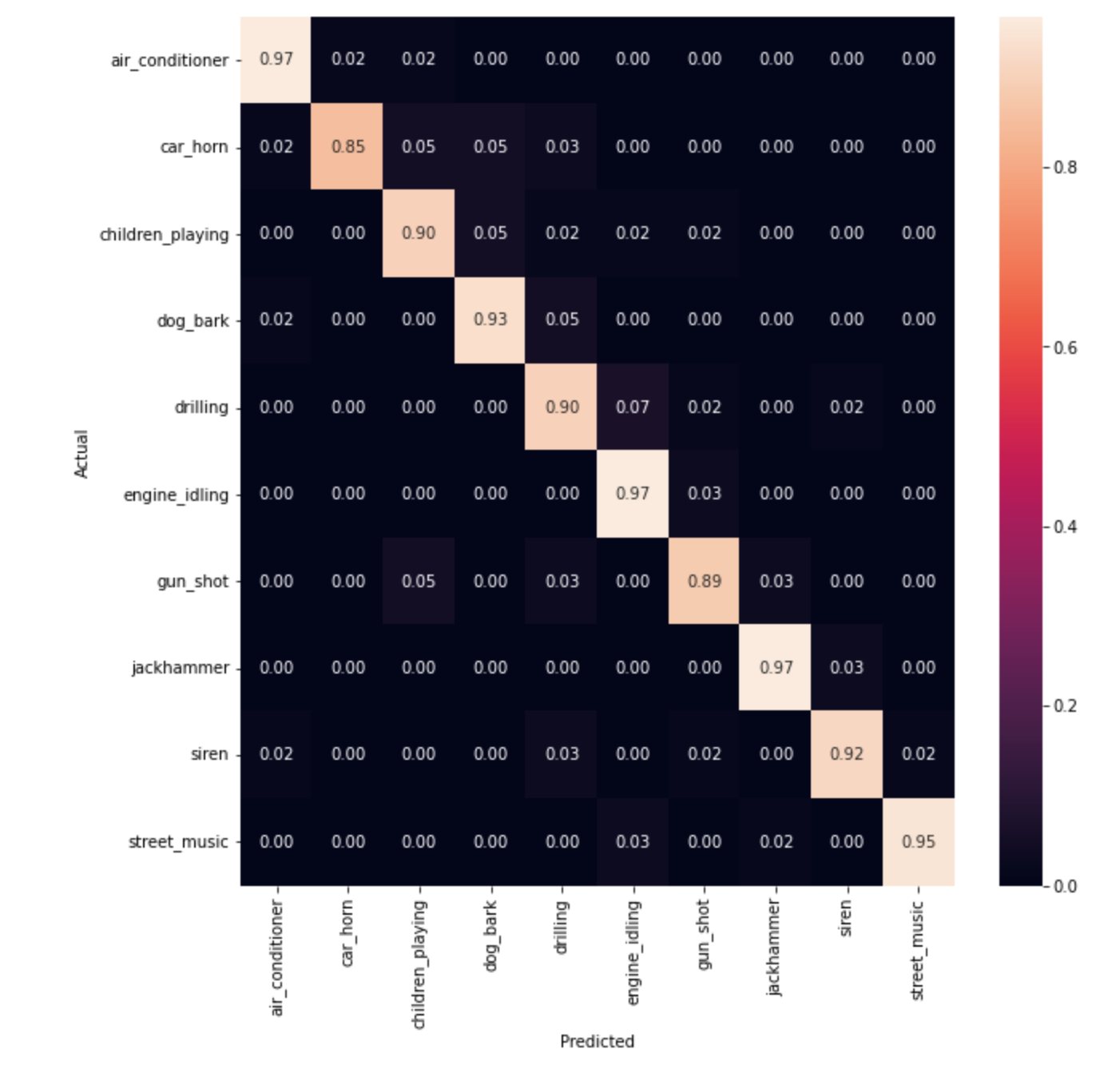}
		\caption{Confusion matrix of evaluation results  without mini-batch balancing.}
	\end{figure}
When mini-batch balancing is applied during training, it improves the  classification accuracy of car horn and gunshot classes  to 91\%  and 93\%  respectively,  justifying the  need for  mini-batch balancing during training to boost the number of instances of the minor classes seen by the learner during training hence improving its ability to generalise well for these classes.
	\begin{figure}[H]
	\centering
\includegraphics[scale=0.4,angle=0]{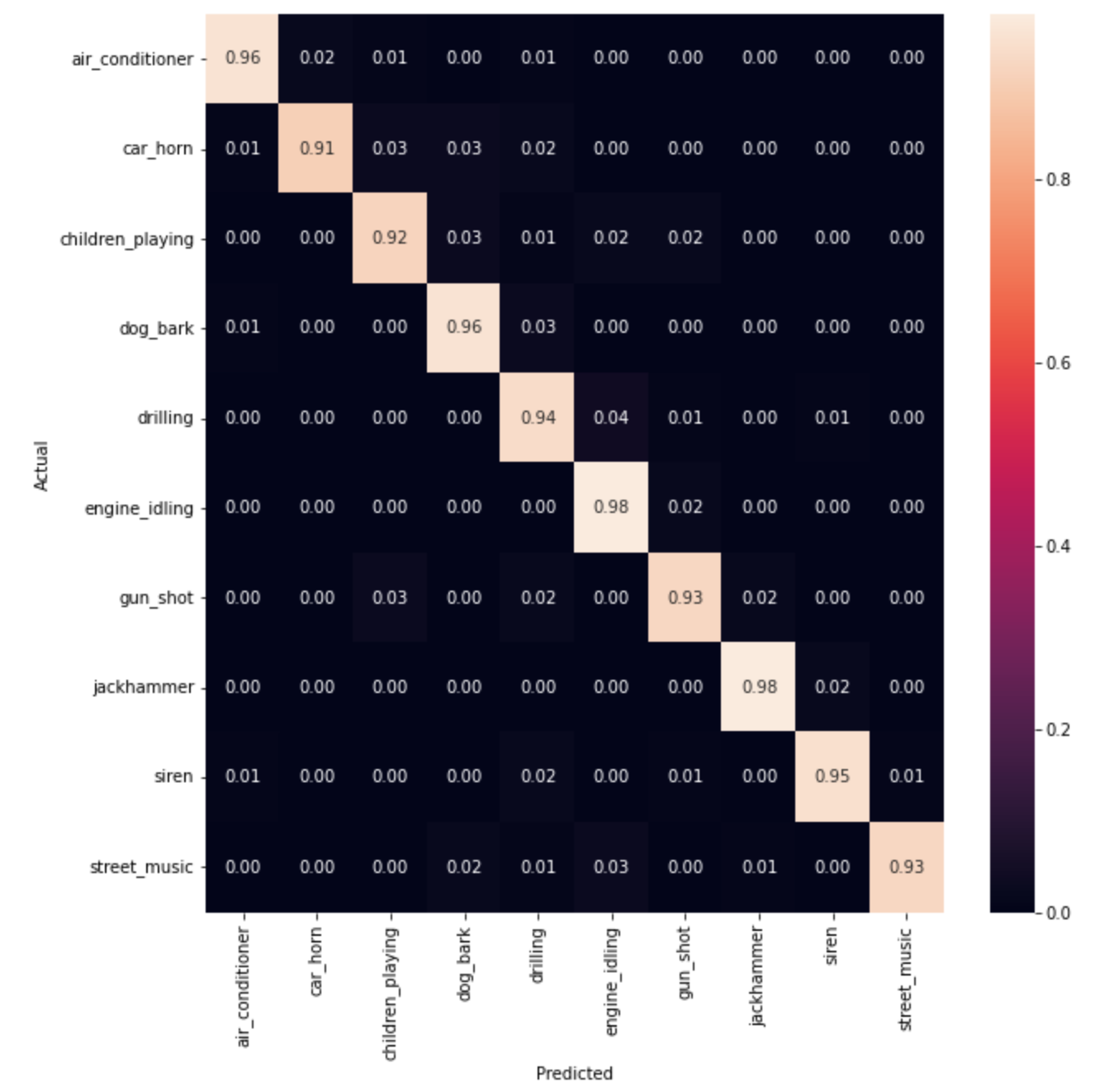}
		\caption{Confusion matrix of evaluation results  with mini-batch balancing.}
	\end{figure}
\subsection{ Comparison to other state of the art tools}
Finally, we compared the results of the proposed technique in classifying environmental sounds and the existing state of the art tools for environmental sound classification. We compare the tools on both the ESC-50 and the UrbanSound8K dataset. The results are shown in table 6. On the ESC-50 the three configurations proposed in this work outperforms the existing methods. Contrastive (raw waveform) the least performing technique proposed outperforms the best performing tool TFCNN by a 10\% margin. This  demonstrates the robustness of the contrastive learning method and the ability of the 1D CNN to capture most of the audio features. On the UrbanSound8K dataset  Contrastive (raw waveform) achieves an accuracy of 95.1\% which is 2.1\% lower than TSCNN the highest performing tool on this dataset. Contrastive(fusion)  method outperforms all the existing techniques in both ESC-50 and UrbanSound8K. It records 12.08\% increase as compared to the best performing tool TFCNN on the $ESC-50$ dataset. On the UrbanSound8K it achieves an increase of 0.9\% as compared to TSCNN is the best performing tool on this dataset.
\begin{table}
\centering
\caption{ Effects of varying CNN depth on the performance of the proposed model}
 \begin{tabular}{ |c | c | c|}
    \hline
   \textbf{Tool} & \textbf{ESC-50}& \textbf{UrbanSound8K}\\ \hline
   Piczak-CNN \cite{Piczac100}& 65.0\%&73.7\% \\ \hline
   TFCNN \cite{TFCNN100} & 84	.4\%&93.1\% \\\hline
   Pyramid CNN\cite{pyramid100} & 81.4\%&78.1\% \\\hline
    SB\_CNN \cite{Salamon2017}& -&79.0\%\\\hline
     SoundNet \cite{Aytar2016}& 74.2\%&-\\\hline
     DS-CNN\cite{Li2018} & 82.8\%&92.2\%\\\hline
      TSCNN \cite{Su2019} &-&97.2\%\\\hline
         Contrastive (raw waveform) & 94.42& 95.1 \\\hline
     Contrastive (spectrogram patches) & 95.3 & 96.1\\\hline
          Contrastive (fusion)  & 97.2 & 98.1\\\hline
               \end{tabular}
                  \end{table}
      \section{Conclusion}
      This work proposes the use of self-supervised contrastive learning to train a learner which can  extract features from the environmental sounds. We exploit a shallow 1D CNN network to extract features of a given audio. We examine the effect of type of input on the quality of representations generated by the learner. Further, we examine the effect of fusing representations from two types of input. The mini-batch balancing is also performed to improve the generalisation of the minority classes by the learner. Overall the proposed technique records a significant improvement in accuracy in the classification task as compared to the existing methods. This demonstrated the robustness of the contrastive learning method and the ability of the 1D CNN to extract the features of waveform.
\bibliographystyle{unsrtnat}
\bibliography{MyCollection}   

\end{document}